\documentclass[aps,showpacs,pra,twocolumn,floatfix]{revtex4-2}
\usepackage[T1]{fontenc}
\usepackage[margin=0.7in,columnsep=0.25in]{geometry}
\usepackage{amsmath,amssymb,amsthm,graphicx,float,lmodern,microtype}
\usepackage{physics2,ragged2e,booktabs}
\usephysicsmodule{ab,braket,op.legacy}
\usepackage[dvipsnames]{xcolor}
\usepackage{tikz}
\usetikzlibrary{arrows.meta,positioning,calc}
\usepackage[hidelinks]{hyperref}
\hypersetup{pdftitle={Explicit absolutely maximally entangled states in dimensions twice a prime power},
 pdfauthor={Palash Pandya}}
\usepackage{needspace}
\newcommand{\AME}{\operatorname{AME}}
\newcommand{\wt}{\operatorname{wt}}
\newcommand{\F}{\mathbb F}
\newcommand{\Id}{\mathbb I}
\newcommand{\bl}{\boldsymbol l}
\newcommand{\file}[1]{\nolinkurl{#1}}
\newtheorem{theorem}{Theorem}[section]
\newtheorem{proposition}[theorem]{Proposition}
\newtheorem{corollary}[theorem]{Corollary}
\newtheorem{lemma}[theorem]{Lemma}

\begin{document}
\title{Explicit Absolutely Maximally Entangled states in dimensions twice a prime power}
\author{Palash Pandya}
\affiliation{Center for Theoretical Physics, Polish Academy of Sciences, Aleja Lotnik\'{o}w 32/46, 02-668 Warsaw, Poland}
\affiliation{Center for Quantum-Enabled Computing, Center for Theoretical Physics, Polish Academy of Sciences, Aleja Lotnik\'{o}w 32/46, 02-668 Warsaw, Poland}
\date{\today}

\begin{abstract}
Absolutely maximally entangled (AME) states are pure multipartite states with
maximally mixed reductions on every set of at most half the parties.
We construct explicit families in local dimension $2q$, twice an odd prime
power $q$, using an established coupling of qubits and $q$-level systems.
The $q$-level states remove interference between branches, so the qubit
states need maximally mixed reductions only on average. This gives
AME$(9,2q)$ and AME$(11,2q)$ for every odd prime power $q\geq5$,
AME$(13,2q)$ for $q\geq17$, AME$(15,2q)$ for $q\geq9$, and
AME$(17,2q)$ for $q\geq67$. For every odd $N\geq3$, the same
coupling gives AME$(N,2q)$ under the sufficient condition
$q\geq\max\{N,2^{N-2}\}$ on the odd prime power $q$. Within this
coupling, the smallest such $q$ must grow exponentially with $N$.
Combined with known inputs and products, these
families complete the nine-party classification: for integer $d\geq2$,
an AME$(9,d)$ state exists if and only if $d\geq3$. Choosing the branch
pairing also gives AME$(15,2p)$ states with a cyclic shift of the parties
compensated by local unitaries, for every prime $p\equiv1\pmod{15}$.
\end{abstract}
\maketitle

\section{Introduction}\label{sec:intro}

A pure state of $N$ parties of local dimension $d$ is absolutely maximally
entangled~\cite{scott,helwig2012}, written AME$(N,d)$, if every reduction to half the parties is
maximally mixed. More precisely, for a retained subset $A$, its complement
$A^c$ and the identity $\Id_A$ on its Hilbert space, the condition is
\begin{equation}
 \rho_A=\Tr_{A^c}\,\ketbra{\psi}{\psi}
       =\frac{\Id_A}{d^k},\qquad |A|=k=\lfloor N/2\rfloor.
 \label{eq:ame}
\end{equation}
Tracing out further parties preserves maximal mixedness, so every smaller
reduction is maximally mixed as well. More generally, an $N$-party density
matrix, pure or mixed, is called \emph{$k$-uniform} when all its $k$-party
reductions are maximally mixed~\cite{scott,goyeneche2014,klobus2019mixed}. An AME$(N,d)$ state is thus a pure
$\lfloor N/2\rfloor$-uniform state, the highest uniformity a pure state can
have. Purity is what makes the condition demanding: the maximally mixed
state is trivially $k$-uniform for every $k$, whereas in a pure AME state
every group of at most half the parties sees a completely random state,
so all information sits in the correlations. This property connects multipartite entanglement with quantum
error correction and secret sharing~\cite{helwig2012,helwigcui,hubergrassl,raissi2018}.
Reference~\cite{amereview} reviews the known constructions and open cases.

For a composite local dimension $d=mn$, the tensor product of an
AME$(N,m)$ and an AME$(N,n)$ state is AME$(N,mn)$, with each party
holding one subsystem of each~\cite{helwig2013graph,shi2021masking}. For $d=2q$ with integer $q\geq2$, this would require an AME
state of qubits. For $N\geq2$, such states exist only at
$N=2,3,5,6$~\cite{higuchi2000,rains,scott,hgs7}. The product route therefore cannot supply
the factor two for any of the party counts considered here. The same
obstacle arises for AME$(4,6)$, which was found by other
means~\cite{rather2022}. Stabilizer states cannot avoid it either. For
odd $q$, a stabilizer AME$(N,2q)$ state in the cyclic Weyl convention
would give a stabilizer AME$(N,2)$
state~\cite[Definition~2 and Theorem~2]{cha2026}. Thus the families
constructed here are not stabilizer states in that convention. A
separate obstruction excludes canonical graph-state AME constructions in
even local dimensions when the number of parties is divisible by
four~\cite{wojcik2026}.

A superposition of products avoids this obstacle. We pair several qubit
states with suitably chosen $q$-level states. The latter remove the
interference between different terms, leaving the qubit states to have
maximally mixed reductions only on average. They come from projecting one
extra party of an AME state, a standard step in quantum
coding~\cite{rains1998we,hubergrassl}. As in Ref.~\cite{alsina2021}, we
call this state a parent. The construction thus
asks for a qubit mixture and a parent, rather than an AME state of qubits.

This mechanism has precedents. Earlier constructions combined a
classical code with a basis of uniform quantum states~\cite{raissi2020},
and several recent AME states can be read as superpositions of
combinatorial designs or stabilizer states~\cite{amereview}.
Shi, Zhang, Zhao, and Li construct
AME$(7,2q)$ states for every odd integer $q\geq3$ by pairing two Steane
logical states with odd-dimensional graph states~\cite{shi7}. Their two
factors share the cancellation on different sets of cuts. In the
AME$(9,10)$ state of the Atlas Lean library~\cite{atlas}, proved using
the h1 harness~\cite[Table~1 and App.~D.1]{h1}, and in Kitamura's
AME$(11,10)$~\cite{kitamura},
the five-level factor alone cancels interference on every required cut.
We extend this latter construction, not the coupling principle itself.

Our contribution is to give explicit infinite families with concrete
thresholds and short specifications. The field constructions prove the
infinite ranges analytically. Exact arithmetic verifies the fixed inputs,
such as the qubit mixtures and the parents for smaller fields.
Several of these ingredients are known: Atlas and Kitamura supply the
nine- and eleven-qubit states, classical codes supply the other qubit
groups. Several parents come from Bevins--Bidav and
Hontarenko--\.{Z}yczkowski~\cite{bb,hz}. Their parent results do not
use the coupling.
The $q=5$ members at nine and eleven parties are the known AME$(9,10)$
and AME$(11,10)$ states.

To our knowledge, explicit families at $N=13,15,17$, the $N=9,11$
families beyond those $q=5$ members, and the classification of AME$(9,d)$
have not appeared before. For each fixed $N$, existence for all sufficiently large integer local
dimensions $d$ follows non-constructively from Keevash's existence of
designs~\cite[Definition~1.6 and Theorem~1.7]{keevash}. Its partite form
gives an orthogonal array of index one and strength $\lfloor N/2\rfloor$
over every sufficiently large alphabet, and the code construction of Helwig and
Cui~\cite[Theorem~2]{helwigcui} turns this array into an AME state. Here the states and their thresholds are explicit.
The thresholds are sufficient and do not assert nonexistence below them.

The same mechanism works for every odd number of parties. Using the
known qubit mixtures described in Appendix~\ref{subsec:odd-mixture},
we obtain AME$(N,2q)$ for every odd $N\geq3$ and every odd prime power
$q\geq\max\{N,2^{N-2}\}$. The sharper thresholds of
Theorem~\ref{thm:main} use fewer branches.

For odd prime powers $q\geq N$, the extended Cauchy construction
supplies the parent~\cite{raissi2018}. The all-cut coupling then works exactly when an
$N$-qubit mixture of rank at most $q$ has maximally mixed reductions
on every set of $(N-1)/2$ parties (Theorem~\ref{thm:odd-rank}).
The packing argument bounds how this rank requirement grows, without
determining an optimal rate.

Section~\ref{sec:results} develops the coupling, constructs the families,
and draws three consequences: the nine-party classification, a cyclic
fifteen-party family, and bounds on how far this coupling can go.
Section~\ref{sec:discussion} closes with the remaining questions.
The appendices give the constructions and proofs behind the ingredients,
and the companion repository~\cite{repo} supplies the finite data and exact
checks.

\section{Construction and consequences}\label{sec:results}
\subsection{How the coupling works}\label{sec:coupling}

Set $N=2k+1$. The coupling itself uses only a local dimension
$q\geq2$, which may be any integer. In computational bases, the local
identification is
\begin{equation}
 \ket*{b}_2\ket*{j}_q\longleftrightarrow \ket*{qb+j}_{2q},
 \quad b\in\{0,1\},\quad 0\leq j<q.
 \label{eq:digits}
\end{equation}
At every site the qubit and the $q$-level system belong to the same
party, so a reduction to a set $A$ keeps both factors of every party
in $A$.

We now replace the product by a superposition. A branch pairs a
normalized state $\ket{u_a}$ of the $N$ qubits with a state $\ket{v_a}$
of the $N$ $q$-level systems. Let $K$ be the number of branches, indexed
by $a=0,\ldots,K-1$. Reducing the superposition across a cut $A|A^c$
gives a diagonal term from each branch and interference terms between
distinct branches. For two labels $a,b$, the corresponding term contains
$\Tr_{A^c}\ketbra{u_a}{u_b}\otimes
\Tr_{A^c}\ketbra{v_a}{v_b}$.

We choose the $q$-level states so that their reduced outer products,
called cross blocks, vanish when $a\neq b$ and are maximally mixed when
$a=b$. The qubit mixture then needs only maximally mixed reductions on
average. In the proposition, $\delta_{ab}$ is one for equal labels and
zero otherwise, and $\lambda_a$ are the probabilities of the branches.

\begin{proposition}[Coupling]\label{prop:coupling}
For $a=0,\ldots,K-1$, let $\ket{u_a}$ be normalized $N$-qubit states,
$\ket{v_a}\in(\mathbb C^q)^{\otimes N}$, and
$\lambda_a\geq0$ with $\sum_a\lambda_a=1$. Let
$\rho_{\rm bin}=\sum_a\lambda_a\ketbra{u_a}{u_a}$ be the qubit mixture.
Suppose that every set $A$ of $|A|=k$ parties satisfies
\begin{align}
 \Tr_{A^c}\rho_{\rm bin}&=\Id_A/2^k,
 \label{eq:binary-average}\\
 \Tr_{A^c}\,\ketbra{v_a}{v_b}&=\delta_{ab}\Id_A/q^k.
 \label{eq:erasure}
\end{align}
Then the state
\begin{equation}
 \ket{\Psi}=\sum_a\sqrt{\lambda_a}\ket{u_a}\otimes\ket{v_a}
 \label{eq:coupled}
\end{equation}
is normalized and AME$(N,2q)$ under Eq.~\eqref{eq:digits}.
The qubit states $\ket{u_a}$ need not be orthogonal.
\end{proposition}

\begin{proof}
Tracing Eq.~\eqref{eq:erasure} gives
$\braket{v_b}{v_a}=\delta_{ab}$, so the norm of the coherent sum is
$\sum_a\lambda_a=1$. Its reduced cross terms with $a\ne b$ vanish. Thus
\begin{equation}
 \rho_A=\left(\Tr_{A^c}\rho_{\rm bin}\right)\otimes\frac{\Id_A}{q^k}
       =\frac{\Id_A}{(2q)^k}.
 \label{eq:coupled-reduction}
\end{equation}
\end{proof}

The proof uses no uniformity of the individual qubit states. This is why
they can be used even when no pure AME state of qubits exists. The
$q$-level requirement is stronger than orthogonality: an outer product
between orthogonal states has zero trace, but can still have a nonzero
partial trace. We require that partial trace to vanish on every set of
$k$ parties.

Equivalently, the $q$-level states span a $k$-uniform subspace: every
normalized superposition in it has maximally mixed $k$-party
reductions~\cite{hubergrassl,antipin}. This is different from requiring
only the normalized projector onto a subspace to be $k$-uniform, which
is enough on the qubit side.

A parent supplies the stronger condition automatically. Let
$\ket{\Phi}$ be an AME$(N+1,q)$ state and call its extra party the
reference. Projecting the reference onto different vectors of an
orthonormal basis gives $q$ normalized states with exactly the diagonal
and cross-block reductions required above. We can select any $K\leq q$
of them. This is a known projection step for quantum
codes~\cite[Theorem~19]{rains1998we}. Appendix~\ref{app:parent} proves
it for every integer $q\geq2$.

\begin{figure}[t]
\centering
\begin{tikzpicture}[box/.style={draw,rounded corners=2pt,align=center,
 inner sep=4pt,font=\small,text width=3.2cm},
 arr/.style={-{Latex[length=1.8mm]},thick}]
\node[box] (b) at (0,0) {Qubit mixture $\rho_{\rm bin}$\\$k$-uniform, Eq.~(\ref{eq:binary-average})};
\node[box] (p) at (3.6,0) {Parent $\ket{\Phi}$\\AME$(N+1,q)$};
\node[box] (v) at (3.6,-1.3) {Project reference\\cross blocks vanish};
\node[box,text width=6.8cm] (s) at (1.8,-2.6)
 {$\sum_a\sqrt{\lambda_a}\ket{u_a}\ket{v_a}$\\
 every $k$-party reduction is $\Id_A/(2q)^k$};
\draw[arr] (p)--(v);\draw[arr] (b.south)--($(s.north)+(-1.8,0)$);
\draw[arr] (v.south)--($(s.north)+(1.8,0)$);
\end{tikzpicture}
\caption{Coupling on $N=2k+1$ parties. The qubit mixture
$\rho_{\rm bin}=\sum_a\lambda_a\ketbra{u_a}{u_a}$ must have maximally mixed
reductions on every set of $k$ parties. Projecting the reference party of the
parent $\ket{\Phi}$ gives the $q$-level states $\ket{v_a}$, whose cross blocks
vanish (Lemma~\ref{lem:parent}). Pairing the qubit and the $q$-level system of
each party by Eq.~\eqref{eq:digits} gives local dimension $2q$.}
\label{fig:construction}
\end{figure}
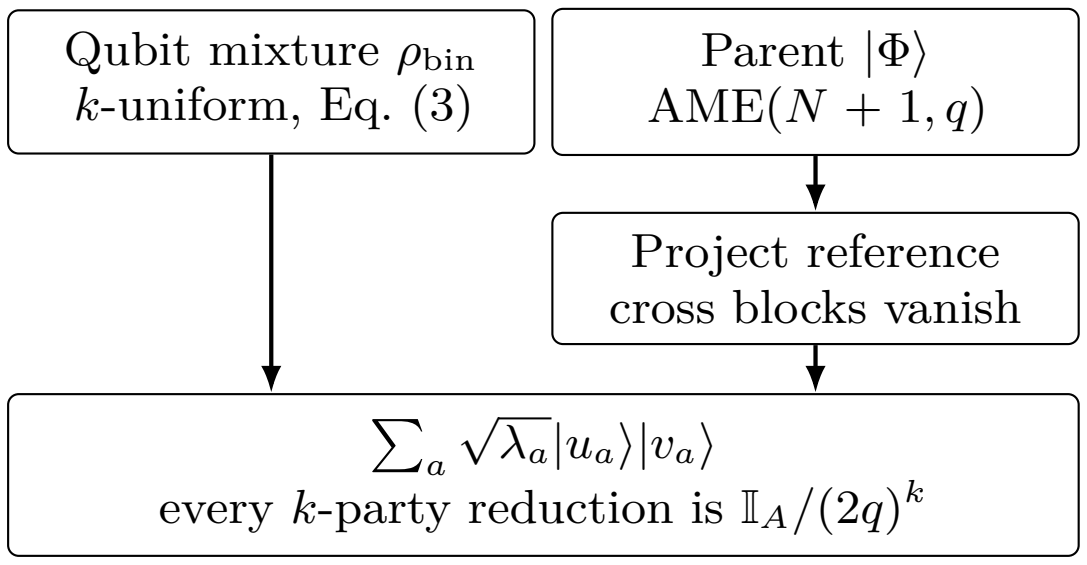

Figure~\ref{fig:construction} summarizes the two tasks. We must find a
$k$-uniform qubit mixture and a parent with enough projected states.
The first task determines how many branches we need. The second determines
which local dimensions can supply them.

\subsection{Explicit families}\label{sec:ingredients}
The qubit mixtures stay fixed while the field size varies. For the family
statements, write $q=p^e$, where $p$ is an odd prime and $e\geq1$ is an
integer, and let $\F_q$ be the field of $q$ elements. Statements written
with $p$ instead of $q$ concern primes only.

The coupling gives a baseline for every odd number of parties. Using a
known family of qubit mixtures, we obtain the following explicit
sufficient range.

\begin{corollary}[Any odd number of parties]\label{cor:odd-N}
Let $N=2k+1\geq3$, with $k$ an integer, and set
$B_N=\max\{N,2^{N-2}\}$. For every odd prime power $q\geq B_N$,
an AME$(N,2q)$ state exists through the coupling of
Proposition~\ref{prop:coupling}.
\end{corollary}

The mixture in Corollary~\ref{cor:odd-N} uses $2^{N-2}$ branches, as
Appendix~\ref{subsec:odd-mixture} explains. More economical ingredients
give much smaller dimensions at the five party counts below. Here
$q_N$ denotes the sufficient threshold for each family.

\begin{theorem}\label{thm:main}
For each $N$ below and every odd prime power $q\geq q_N$, an
AME$(N,2q)$ state exists.
\begin{center}\upshape
\setlength{\tabcolsep}{9pt}
\begin{tabular}{lccccc}
\toprule
$N$ & 9 & 11 & 13 & 15 & 17\\
\midrule
$q_N$ & 5 & 5 & 17 & 9 & 67\\
$2q_N$ & 10 & 10 & 34 & 18 & 134\\
\bottomrule
\end{tabular}
\end{center}
\end{theorem}

The qubit mixtures use $K=4,2,16,8,64$ branches for
$N=9,11,13,15,17$, respectively. In each case we take equal weights on
an orthonormal basis of a subspace. If $P$ is its projector and has rank
$K$, the mixture is $\rho_{\rm bin}=P/K$. It is the reductions of this
mixture that must be maximally mixed, not those of every vector in the
subspace. At $N=9,11$ we use the known Atlas and Kitamura mixtures.
For $N=13,15,17$ we construct the projectors from commuting Pauli
operators, products of the identity and the three single-qubit Pauli
matrices. Appendix~\ref{app:qubit} gives the groups, proves their
uniformity and writes their individual branches explicitly.

With these mixtures fixed, the ordinary Cauchy construction supplies a
parent for every prime power $q\geq N+1$. Its extended form also covers
$q=N$ when $N$ is a prime power (Lemma~\ref{lem:cauchy}). These are
graph-state forms of the Cauchy-matrix MDS construction of
Raissi et al.~\cite{raissi2018}.
 A second requirement is $q\geq K$, since the
parent has only $q$ projected states. At $N=13,17$ these two requirements
give the thresholds in the theorem. At $N=9,11,15$, separate parents
cover the smaller fields down to $q=5,5,9$. Appendix~\ref{app:parents}
gives the field construction, the finite parents and the full proof of
Theorem~\ref{thm:main}. Table~\ref{tab:families} there records all five
sets of ingredients and their provenance.

For example, the eight qubit states at $N=15$ can be paired with eight
projected states of the AME$(16,13)$ parent to give
\begin{equation}
 \ket{\Psi_{15,26}}=8^{-1/2}\sum_{a=0}^7\,\ket{u_a}\ket{v_a}.
 \label{eq:example}
\end{equation}
On any set of seven parties, the qubit average reduces to $\Id_A/2^7$,
each $q$-level diagonal block to $\Id_A/13^7$, and the cross blocks vanish.
The reduction of $\ket{\Psi_{15,26}}$ is therefore $\Id_A/26^7$.

The state is explicit without storing its full tensor. A qubit graph
with ordered branch labels and a parent graph with ordered reference
values specify every amplitude as a sum of $K$ terms.
Appendix~\ref{subsec:assembly} gives the formulas and basis conventions.
Changing the pairing preserves the AME property but can change the
symmetry, a freedom we use in Sec.~\ref{sec:symmetry}.

First we use the
nine-party family to settle the local-dimension question.

\subsection{Which nine-party dimensions are possible}\label{sec:nine}

Combined with the product rule of the introduction, the $N=9$ family
settles which local dimensions admit an AME state of nine parties. For local dimensions $d_1,d_2$, the
product of AME$(9,d_1)$ and AME$(9,d_2)$, with each party holding one
subsystem of each, is AME$(9,d_1d_2)$, because its four-party reductions are
$\Id_A/d_1^4\otimes\Id_A/d_2^4$. The difficulty is a factor two, since no
AME$(9,2)$ state exists.

\begin{corollary}\label{cor:nine}
For integer local dimensions $d\geq2$, an AME$(9,d)$ state exists
if and only if $d\geq3$.
\end{corollary}

\begin{proof}
Appendix~\ref{app:nine} supplies AME$(9,q)$ for every prime power
$q\geq3$, including even $q$, and describes Kitamura's external
AME$(9,6)$ input. These are the ingredients for the product argument.

If $d$ is odd, factor it into odd prime powers and apply the product
rule. If $4\mid d$, write $d=2^\nu s$ with $\nu\geq2$ and $s$ odd.
The prime-power result handles $2^\nu$ and the odd case handles $s>1$.
For $d=2s$ with $s\geq3$ odd, use Kitamura's AME$(9,6)$
\cite{kitamura96} when $3\mid s$, with residual factor $s/3$.
Otherwise select a prime divisor $p\geq5$ of $s$, use
Theorem~\ref{thm:main}'s AME$(9,2p)$, and multiply by the odd factor
$s/p$. A residual factor one is omitted. Finally Rains' shadow bound
excludes AME$(9,2)$~\cite[Theorem~14]{rains}.
\end{proof}

Products of the known AME$(9,6)$ and AME$(9,10)$ states with odd
prime-power factors already cover every $d=2s$ with $s$ odd and
$3\mid s$ or $5\mid s$. The new family is needed when every prime factor of $s$ is at
least seven, starting at $d=14$. For comparison, the existence question
is settled for all local dimensions up to seven parties, while eight
parties remain open, starting with AME$(8,4)$~\cite{shi7}.

The AME$(9,6)$ input is essential to this argument when the odd factor
has no prime divisor other than three. Our all-cut coupling cannot
supply it: the rank bound in Sec.~\ref{sec:limits} excludes $q=3$.
Kitamura's state instead lets the two factors share the cancellation,
as explained in Appendix~\ref{subsec:external}.

\subsection{Choosing a cyclic symmetry}\label{sec:symmetry}
The AME property leaves freedom in how we pair the branches. We now use
it to arrange a symmetry. Here $q=p$ is prime. A compensated party
symmetry is a permutation that, followed by a product of local unitaries,
returns the state up to a phase. Our local corrections act separately on each
qubit and each $p$-level system. Their explicit form is given
in Appendix~\ref{app:cyclic}.

The parent gives more freedom than selecting its projected basis states.
Any orthonormal set of their superpositions obeys the same reduction
conditions. We can choose those superpositions to match the way a
corrected cyclic shift acts on the qubit branches. Both factors undergo
the same permutation of physical parties. Their actions within the branch
spaces must also match through the pairing.

\begin{theorem}[Cyclic family]\label{thm:c15}
For every prime $p\equiv1\pmod{15}$ there is an AME$(15,2p)$ state
with a compensated cyclic permutation of all fifteen parties.
The first prime in this family is $31$.
\end{theorem}

For the fifteen-party construction, the physical permutation has order
fifteen, while its actions within both branch spaces have order three.
The qubit action is fixed by the eight branches. We choose a parent whose
action has enough eigenvectors of each required eigenvalue, then pair
the matching vectors. Appendix~\ref{app:cyclic} gives the corrected
physical shifts, derives the matching condition and proves the theorem
for every stated prime. It also gives a definite pairing at $p=31$.

Theorem~\ref{thm:c15} gives one cyclic symmetry. We do not determine the
full symmetry group, which could include local unitaries that mix the
qubit and the $p$-level system.

\subsection{How far this coupling can go}\label{sec:limits}
Could different ingredients lower the thresholds? The $q$-level side can
supply at most $q$ branches satisfying Eq.~\eqref{eq:erasure}, even when
they do not come from a parent. This capacity bound holds for every
integer $q\geq2$, as Appendix~\ref{subsec:capacity} proves.

The qubit side gives the other limit. A mixture of $K$ pure states has
rank at most $K$, so a lower bound on its rank is also a lower bound on
the number of branches. This must allow unequal weights and
nonorthogonal branches, not only normalized projectors.

\paragraph*{The general rank criterion.}
For an odd number $N=2k+1\geq3$ of parties, with integer $k\geq1$,
let $\mu_N$ be the minimum rank of a $k$-uniform density matrix on
$N$ qubits. This minimum is attained because the maximally mixed state
qualifies and the possible ranks are positive integers. It is not
restricted to projectors or equal spectra.

\begin{theorem}[General rank criterion]\label{thm:odd-rank}
Let $q$ be an odd prime power with $q\geq N$. The hypotheses of
Proposition~\ref{prop:coupling} can be satisfied at local dimension
$2q$ if and only if $\mu_N\leq q$. In the sufficient direction, the
$q$-level states can be projected from an extended-Cauchy
AME$(N+1,q)$ parent, with $K=\mu_N$ branches.
\end{theorem}

Appendix~\ref{subsec:odd-criterion} proves this criterion by diagonalizing
the qubit mixture and projecting an extended-Cauchy parent. Below $q=N$,
the rank inequality alone does not provide a parent or another suitable
set of erasing branches.

The criterion separates the rank requirement from the parent problem
once $q\geq N$. Special parents can also work below this range, as in
several of our small-field examples. We now give finite rank bounds
that constrain the mixtures.

\begin{proposition}[Qubit rank bounds]\label{prop:rank}
Every $\lfloor N/2\rfloor$-uniform density matrix of $N$ qubits has
rank at least $4,9,8,23$ for $N=9,13,15,17$, respectively, without any
assumption on its spectrum.
\end{proposition}

Appendix~\ref{subsec:rankproof} proves these bounds using exact
inequalities for arbitrary spectra. Together with the capacity bound,
they require $\rank\rho_{\rm bin}\leq K\leq q$. These are limits of the
coupling in Proposition~\ref{prop:coupling}, not of AME states in general.

At $N=9$, rank at least four exceeds $q=3$, so this coupling cannot give
AME$(9,6)$. This explains both the threshold $q=5$ and the need for
Kitamura's input in Sec.~\ref{sec:nine}. Appendix~\ref{subsec:external}
describes how that state avoids the all-cut requirement.

Among odd prime powers, the rank-eight bound at $N=15$ excludes every
$q<9$. Our construction reaches every $q\geq9$, so its threshold cannot
be lowered within this coupling.

At $N=13$, the rank-nine bound excludes $q\leq7$, while $q=9,11,13$
remain open. A stabilizer projector has rank a power of two, so its rank
must be at least sixteen. This explains the branch count used in our
$N=13$ family. The rank-eight projector at $N=15$ already attains the
general rank bound.

At $N=17$, rank at least twenty-three excludes $q\leq19$. The odd prime
powers $23\leq q\leq61$, including $25,27,49$, remain open.

For AME$(13,10)$ the rank-nine bound excludes both this coupling and
its reversed assignment of the two factors. Neither exclusion applies
to other ways of cancelling interference. In contrast, a six-uniform
qubit mixture of rank between nine and thirteen would suffice for
AME$(13,26)$ with the existing AME$(14,13)$ parent. We have not found
such a mixture. Appendix~\ref{subsec:consequences} gives both arguments.

Within this all-cut coupling, the required local dimension grows at least
exponentially with odd $N$ by the rank bound in
Appendix~\ref{subsec:odd-growth}.

\section{Conclusions and future work}\label{sec:discussion}
We have constructed five explicit infinite families of AME states in
local dimensions twice an odd prime power. At these party counts, a
pure qubit AME state is unavailable, so taking a product with an
odd-dimensional AME state cannot supply the factor two. Our coupling
uses a qubit mixture with maximally mixed small reductions and a parent
whose projected branches remove interference. Reusing the qubit
ingredients as the field size varies gives the sufficient ranges in
Theorem~\ref{thm:main}. Analytic field constructions establish the
infinite ranges, while the companion repository~\cite{repo} supplies
exact certificates and reconstruction code for the fixed inputs.

The same coupling extends beyond these five party counts. A known qubit
mixture and an extended-Cauchy parent give AME$(N,2q)$ for every odd
$N\geq3$ whenever the odd prime power $q$ satisfies
$q\geq\max\{N,2^{N-2}\}$. The five explicit families improve on this
general bound through more economical mixtures. In the parent regime
$q\geq N$, Theorem~\ref{thm:odd-rank} characterizes feasibility of the
all-cut coupling by the minimum rank of a qubit mixture with the required
uniform reductions. This reduces the remaining existence question within
that regime to constructing suitable qubit density matrices.

At nine parties, our family combines with prime-power constructions,
products and Kitamura's AME$(9,6)$ state~\cite{kitamura96} to give
AME$(9,d)$ for every integer $d\geq3$. Together with nonexistence at
$d=2$, this completes the integer-dimension classification in
Corollary~\ref{cor:nine}. The external dimension-six state is an
essential input to this conclusion.
At fifteen parties, the freedom to pair the branches also allows us to
impose a compensated cyclic symmetry involving all parties.
Theorem~\ref{thm:c15} gives such a family for every prime
$p\equiv1\pmod{15}$, starting at $p=31$. Thus the pairing can enforce
symmetry while preserving the AME reductions. The displayed symmetry
includes local corrections; the full symmetry group remains to be
determined.

Two directions could extend these results. Within the all-cut coupling,
lower-rank qubit mixtures could reduce the required local dimensions.
A concrete target is a six-uniform thirteen-qubit mixture of rank between
nine and thirteen: combined with the existing AME$(14,13)$ parent, it
would give AME$(13,26)$. More generally, the gap between the sufficient
dimension bound and the exponential lower bound leaves open the optimal
growth rate within this coupling. Beyond it, allowing the two factors
to share the cancellation, as in Ref.~\cite{shi7}, could reach states
excluded by our all-cut conditions, including potentially AME$(13,10)$.
A further question is which additional symmetries can be imposed through
the branch pairing and how they depend on the chosen ingredients.

\begin{acknowledgments}
This work was supported by the QuantERA II Programme (VERIqTAS project), which has received funding from the European Union's Horizon 2020 research and innovation programme under Grant Agreement No. 101017733. We also acknowledge funding from the European Union's Horizon Europe research and innovation programme under Grant Agreement No. 101080086 (NeQST).
\end{acknowledgments}

\paragraph*{Declaration of AI use}
OpenAI Codex and Anthropic Claude Code were used to assist with
mathematical exploration, literature searches, and code development using Python and Wolfram Mathematica. They were also used for drafting, checking and revising the manuscript. The author supervised the use of these tools and takes
responsibility for the scientific content and final manuscript.

\appendix
\section{Why projecting a parent supplies the branches}\label{app:parent}
The coupling requires the parent branches to have both uniform reductions
and vanishing cross blocks. We prove these properties together by keeping
the reference party when taking a reduction. The projection argument is
standard for quantum codes~\cite[Theorems~19 and~20]{rains1998we}, see
also Ref.~\cite[Theorem~2 and Proposition~7]{hubergrassl}. It also
appears in the information-masking construction of Shi, Li, Chen, and
Zhang~\cite[Theorem~10 and Corollary~12 of arXiv v2]{shi2021masking}.
Throughout this appendix,
$N=2k+1$ and $q\geq2$ is an integer. Label the parent parties
$0,1,\ldots,N$, with party $0$ the reference.

\subsection{Projection and normalization}\label{subsec:projection}

\begin{lemma}[Projected parent]\label{lem:parent}
Let $\ket{\Phi}$ be an AME$(N+1,q)$ state on parties $0,1,\ldots,N$.
Project party $0$ onto a state $\ket{c}$ of an orthonormal basis:
\begin{equation}
 \ket{v_c}=\sqrt q\,(\bra*{c}_0\otimes\Id)\ket{\Phi},
 \qquad c=0,\ldots,q-1.
 \label{eq:conditioning}
\end{equation}
Then the $q$ states $\ket{v_c}$, and hence any $K\leq q$ of them, satisfy
Eq.~\eqref{eq:erasure}.
\end{lemma}

The factor $\sqrt q$ normalizes $\ket{v_c}$, because the reference is
maximally mixed and each outcome $c$ occurs with probability $1/q$.

The reference labels here are integers indexing any orthonormal basis.
Select any $K\leq q$ of them and reindex the states by the branch label
$a=0,\ldots,K-1$. No finite-field structure is needed for this step.

\subsection{The reduced blocks}\label{subsec:parentproof}
To check all pairs of labels at once, expand the parent over the reference.
Its reduction on the reference and $k$ other parties fixes every block.

\begin{proof}
Inserting $\sum_c\ketbra{c}{c}=\Id_0$ on party $0$ and using
Eq.~\eqref{eq:conditioning} gives
$\ket{\Phi}=q^{-1/2}\sum_c\,\ket{c}_0\ket{v_c}$. Keep party $0$ and a set
$A$ of $k$ other parties. Together these are $k+1=(N+1)/2$ parties, half
of the parent, so their reduced state is $\Id/q^{k+1}$. Computed from the
expansion, the same reduced state is
\begin{equation}
 \rho_{0A}=\frac1q\sum_{c,c'}\ketbra{c}{c'}\otimes
           \Tr_{A^c}\,\ketbra{v_c}{v_{c'}}.
 \label{eq:parent-blocks}
\end{equation}
Matching the two expressions block by block gives Eq.~\eqref{eq:erasure}.
\end{proof}

\section{The explicit qubit mixtures}\label{app:qubit}\label{subsec:binary}
The main construction needs a $k$-uniform qubit mixture, where
$k=\lfloor N/2\rfloor$. We describe it by a projector, then choose a
basis to obtain the individual branches. The distinction matters: only
their average must be uniform.

\subsection{Projectors and the weight test}\label{subsec:weights}

In every case we take equal weights on an orthonormal basis of a
subspace. If $P$ projects onto that subspace and has rank $K$, the mixture
is $\rho_{\rm bin}=P/K$, so the rank is also the branch count. At $N=9$
and $11$ such mixtures are known. Atlas uses four branches of the
nine-cycle graph, and Kitamura uses two branches obtained by projecting one
qubit of a five-uniform twelve-qubit graph state~\cite{atlas,kitamura}.
Their averages are four- and five-uniform, respectively, although the
individual branches need not be uniform. For $N=13,15,17$ we build $P$
from commuting Pauli operators, which describe it compactly and make its
reductions easy to check.

Zhang, Pang, Fei, and Zheng also give a four-uniform nine-qubit normalized
projector of rank eight~\cite[Sec.~II, Example~4]{zhang2024mixed}.
The four-branch Atlas mixture used here has rank four~\cite{atlas},
with the matching lower bound proved in Proposition~\ref{prop:rank}.

A Pauli word is a tensor product of $I,X,Y,Z$. Its weight $\wt(E)$ is the
number of nonidentity factors. Let $S_{\rm bin}$ be an abelian group of
Hermitian Pauli operators with $r$ independent generators and
$-\Id\notin S_{\rm bin}$. Its common fixed space has projector and rank
\begin{equation}
 P=2^{-r}\sum_{E\in S_{\rm bin}}E,\qquad K=\Tr P=2^{N-r}.
 \label{eq:projector}
\end{equation}
This normalized-projector construction is used by
K{\l}obus et al.~\cite[Sec.~III, Eqs.~(8)--(12)]{klobus2019mixed}
for $k$-uniform mixed states.
The normalized projector $P/K$ is $k$-uniform exactly when every nonidentity
word in $S_{\rm bin}$ has weight greater than $k$, which is the condition
used in Ref.~\cite[Sec.~III]{klobus2019mixed}. We include the short proof.
Indeed,
$P/K=2^{-N}\sum_E E$, and tracing any nonidentity Pauli on a discarded
site gives zero. Only words supported inside $A$ can remain. Distinct
surviving words cannot cancel: $-\Id\notin S_{\rm bin}$ ensures distinct
Pauli words, which are linearly independent. This gives both directions
of the criterion.

The criterion makes only the average over a basis of the fixed space
$k$-uniform, which is all that Proposition~\ref{prop:coupling} requires.
It does not make every pure state in that space $k$-uniform. It concerns the weights
of the stabilizing operators, not the error-correcting distance of the
subspace. This is why the projector can supply the qubit mixture even
when no pure AME state of qubits exists.

\subsection{The three code constructions}\label{subsec:codes}
We now give the groups for $N=13,15,17$. A four-element field records
Pauli words compactly, so that known classical codes supply the seeds.

To write the stabilizer groups compactly, we use the correspondence
between Pauli words and vectors over the four-element
field~\cite{crss1998}. We label single-qubit Pauli
operators by the elements of the four-element field
$\F_4=\{0,1,\omega_4,\omega_4^2\}$, where $\omega_4^2=\omega_4+1$. The
labels $0,1,\omega_4,\omega_4^2$ stand for $I,X,Z,Y$, with signs ignored.
A Pauli word on $N$ qubits then becomes a vector in $\F_4^N$, and its
weight is the number of nonzero entries. Multiplying two words corresponds
to adding their vectors. A group of Pauli words therefore becomes a set of
vectors closed under addition, called an \emph{additive code}. Two words
$x,y$ commute exactly when $s+s^2=0$, where $s=\sum_i x_iy_i^2$. A code is
\emph{Hermitian self-orthogonal} if $s=0$ for every pair of its vectors.
All its Pauli words then commute, so it describes an abelian group as
required above. We write $[n,\kappa,\delta]_Q$ for a code over $\F_Q$ that
is closed under multiplication by field elements and has length $n$,
dimension $\kappa$, and minimum number of nonzero entries in a nonzero
vector $\delta$.

\begin{proposition}[Qubit ingredients]\label{prop:binary}
For $N=13,15,17$ there are explicit abelian groups $S_{\rm bin}$ of
commuting Pauli words, given in the proof, with
\begin{equation}
 \begin{gathered}
 (N,r,K,w_{\min})=(13,9,16,7),\ (15,12,8,8),\\
 (17,11,64,9),
 \end{gathered}\label{eq:binary-parameters}
\end{equation}
where $w_{\min}$ is the smallest weight of a nonidentity word in the group.
In each case $w_{\min}>\lfloor N/2\rfloor$, so the normalized projector
$P/K$ is $\lfloor N/2\rfloor$-uniform.
\end{proposition}

\begin{proof}
We use seeds of lengths $15$ and $18$ over $\F_4$.
The parameters of the length-$15$ seed occur in
Ref.~\cite{codes15}, while Ref.~\cite{codes17} supplies the
length-$18$ quasicyclic seed. The displayed generators and
their required properties are checked in the companion repository.
Positions are numbered from zero. Hermitian generators with any signs give
a group without $-\Id$, because the generators are independent.

\emph{Length 15.} Write a vector in $\F_4^{15}$ as a polynomial
$c_0+c_1z+\cdots+c_{14}z^{14}$. The code is then described by the single
polynomial
\begin{equation}
 \begin{split}
 g(z)={}&1+\omega_4^2z+z^2+z^4+\omega_4^2z^5\\
       &+\omega_4z^7+\omega_4z^8+z^9.
 \end{split}\label{eq:seed15}
\end{equation}
As a Pauli word, $g(z)$ is $XYXIXYIZZXIIIII$. Multiplying by $z^j$ modulo
$z^{15}-1$ shifts this word cyclically by $j$ qubits, and multiplying by
$\omega_4$ relabels $X\to Z\to Y\to X$ on every qubit. The group is
generated by the six shifts with $j=0,\ldots,5$ and their six relabelled
versions. The code is Hermitian self-orthogonal, so these words commute.
The twelve generators are independent, so $r=12$ and $K=2^{15-12}=8$.
Counting the $2^{12}=4096$ words of the group by weight gives one identity
and $405$, $1260$, $1890$ and $540$ words of weight $8$, $10$, $12$ and
$14$, respectively. The smallest nonidentity weight is therefore $8$,
greater than $\lfloor 15/2\rfloor=7$.

\emph{Removing qubits.} To obtain smaller groups from these seeds, we
restrict the allowed operators on a chosen qubit and then delete it.
Keeping only $I$ or $X$ preserves commutation, because the deleted factors
commute, and lowers each weight by at most one. Keeping only $I$ lowers
no weight. The same operations shorten stabilizer codes
in Ref.~\cite[Theorem~6(b),(d)]{crss1998}. Here we track the weights of the
group itself rather than a code distance. In the cases below, the first
restriction removes one generator and the second removes two.

\emph{Length 13.} In the length-15 group, require $I$ on qubit $0$ and $I$
or $X$ on qubit $1$, then delete both. This removes three generators, so
$r=9$ and $K=16$. Weights stay at least $7$, and enumeration shows that
$7>6$ is attained.

\emph{Length 17.} Start from the quasicyclic $[18,6,10]_4$ code listed
in Ref.~\cite{codes17}. It is Hermitian self-orthogonal, as the
certificates below confirm, and has $r=12$ and minimum weight $10$. It is generated by shifting cyclically, within each of three blocks
of length six, the vector whose blocks are, in ascending powers of $z$,
\begin{equation}
 \begin{gathered}
 (\omega_4,0,0,1,0,1),\quad(0,1,\omega_4,\omega_4,\omega_4,\omega_4),\\
 (\omega_4^2,\omega_4^2,1,1,1,1).
 \end{gathered}\label{eq:seed18}
\end{equation}
Requiring $I$ or $X$ on the last qubit and deleting it gives $r=11$,
$K=64$, and weights at least $9>8$.

The generators and weight counts are checked exactly by the certificates
in the companion repository.
\end{proof}

\subsection{Individual branches and their phases}\label{subsec:graphbranches}

The groups above give the required qubit mixtures. To obtain individual branches and their
amplitudes, we now choose a basis of each fixed space. We extend its
commuting group to a maximal commuting group and use local Clifford
operations, which map Pauli operators to Pauli operators, to put it in
graph form~\cite{vandennest2004}. These local changes of basis preserve maximal mixedness of
the reductions. Let $B$ be the resulting symmetric binary graph matrix
with zero diagonal, and let $\bl_a\in\F_2^N$ be the branch labels in a
fixed order. For $b=(b_1,\ldots,b_N)\in\F_2^N$, the amplitudes
$u_a(b)=\braket{b}{u_a}$ are
\begin{equation}
 \begin{split}
 u_a(b)&=2^{-N/2}(-1)^{f_B(b)+\bl_a\cdot b},\\
 f_B(b)&=\sum_{i<j}B_{ij}b_i b_j.
 \end{split}\label{eq:binary-amplitude}
\end{equation}
The labels form a linear space $\mathcal L$ over $\F_2$ of size $K$, and
the branches are orthonormal because distinct labels give orthogonal sign
patterns. The graph state of $B$ is the common $+1$ eigenvector of the $N$
words $g_i=X_i\prod_jZ_j^{B_{ij}}$~\cite{hein2004}. The product $g^x=\prod_ig_i^{x_i}$, for
$x\in\F_2^N$, has expectation $(-1)^{\bl_a\cdot x}$ in branch $a$, so its
average over the branches is
\begin{equation}
 \frac1K\sum_a(-1)^{\bl_a\cdot x}
   =\begin{cases}1,&x\in\mathcal L^\perp,\\0,&x\notin\mathcal L^\perp,
     \end{cases}
 \label{eq:label-average}
\end{equation}
where $\mathcal L^\perp$ is the set of vectors orthogonal to every label.
Averaging thus keeps exactly the products $g^x$ with $x\in\mathcal L^\perp$,
which form the chosen group in graph form. The graph and the label order
also fix the phases of the branches and the signs in the group. This gives
an efficient check of the mixture, which involves only Pauli words and
branch labels rather than all $4^N$ matrix elements.

For the $N=9,11$ ingredients, exact checks confirm that the average sign
is zero for every graph-stabilizer word of weight at most $k$.

These qubit ingredients stay fixed as $q$ varies.

\section{Explicit parents and complete state specifications}\label{app:parents}\label{subsec:parents}
The qubit mixtures of Appendix~\ref{app:qubit} fix the required branch
counts. We construct parents for the fields that can supply them, first
with one formula and then with the finite inputs below its range.
Here $N=2k+1$, and $q=p^e$ is a prime power with $p$ prime and $e\geq1$.
The family theorem uses odd $p$.

\subsection{Graph parents over a field}\label{subsec:fieldgraphs}

We label the basis states of each $q$-level party by field elements,
$\ket{y}$ with $y\in\F_q$. For prime $q$ these are just the integers
modulo $q$. For $q=p^e$ with $e>1$, we fix any one-to-one map
$\iota:\F_q\to\{0,\ldots,q-1\}$ and identify $\ket{y}$ with
$\ket*{\iota(y)}_q$. This only renames the basis. Set $n=N+1=2h$, so
$h=k+1$.
The phases come from the additive character
$\chi(z)=\exp\ab[2\pi i\operatorname{Tr}(z)/p]$ on $\F_q$. Here
$\operatorname{Tr}(z)=z+z^p+\cdots+z^{p^{e-1}}$ is the field trace, which
takes values in $\F_p$ and has nothing to do with the partial trace. For
prime $q$ it is the identity, so $\chi(z)=e^{2\pi iz/q}$. A weighted graph
on the $2h$ parties is a symmetric matrix $\Gamma$ over $\F_q$ with zero
diagonal, where $\Gamma_{ij}$ is the weight of the edge $ij$. For $y=(y_1,\ldots,y_{2h})\in\F_q^{2h}$, set $Q_\Gamma(y)=\sum_{i<j}\Gamma_{ij}y_iy_j$. Its graph
state is
\begin{equation}
 \ket{\Gamma}=q^{-h}\sum_{y\in\F_q^{2h}}
       \chi\bigl(Q_\Gamma(y)\bigr)\ket{y},
 \label{eq:field-graph}
\end{equation}
A \emph{balanced} cut splits the $2h$ parties into two sets of $h$. For
such a cut $A|A^c$, let $\Gamma_{A,A^c}$ be the $h\times h$ submatrix of
$\Gamma$ with rows in $A$ and columns in $A^c$. Its entries are the
weights of the edges that cross the cut. If this submatrix is invertible,
the reduction to $A$
is maximally mixed. For prime $q$ this is Helwig's
criterion~\cite[Theorem~7]{helwig2013graph}, and for qubits the rank of
the crossing block gives the entanglement across the cut~\cite{hein2004}.
The following argument works over any $\F_q$. Split each basis label as
$y=(y_A,y_{A^c})$. Tracing out $A^c$ means summing over
$y_{A^c}\in\F_q^h$. In the matrix element $\bra{y_A}\rho_A\ket*{y'_A}$,
the terms of $Q_\Gamma$ that involve only $y_{A^c}$ cancel, and the terms
that involve only $y_A$ or $y'_A$ give an overall phase. The only terms
left are the ones that couple the two sides. With
$w=(y_A-y'_A)^{\mathsf T}\Gamma_{A,A^c}$,
\begin{equation}
 \bigl|\bra{y_A}\rho_A\ket*{y'_A}\bigr|
   =q^{-2h}\,\Bigl|\sum_{y_{A^c}\in\F_q^h}\chi(w\cdot y_{A^c})\Bigr|.
\end{equation}
The sum equals $q^h$ when $w=0$. When $w\neq0$ it is zero, because
$w\cdot y_{A^c}$ then takes every value in $\F_q$ equally often and the
values of $\chi$ over $\F_q$ add up to zero. If $\Gamma_{A,A^c}$ is
invertible, $w=0$ only when $y_A=y'_A$. So $\rho_A$ is diagonal with every entry
equal to $q^{-h}$. Thus $\ket{\Gamma}$ is AME whenever $\Gamma_{A,A^c}$
is invertible for every balanced cut.

\subsection{One recipe for the large fields}\label{subsec:cauchyproof}

We now choose $\Gamma$ so that this holds for all balanced cuts at once.
Split the parties into two halves of size $h$ and allow edges only between
the halves. Then $\Gamma$ is fixed by the $h\times h$ matrix $C$ of the
weights between the halves. We show below that each $\Gamma_{A,A^c}$ is
built from two square submatrices of $C$. So it suffices that every square
submatrix of $C$ is invertible. Cauchy matrices have exactly this
property~\cite{macwilliamssloane,rothseroussi}. Raissi et al.\ used them
in this way to build AME states from MDS codes~\cite{raissi2018}, and
Lemma~\ref{lem:cauchy} writes such states directly as graph states.

\begin{lemma}[Cauchy parents]\label{lem:cauchy}
Choose $2h$ distinct field points $\xi_1,\ldots,\xi_h,\eta_1,\ldots,\eta_h$.
With
\begin{equation}
 C_{ij}=(\xi_i-\eta_j)^{-1},\qquad
 \Gamma=\begin{pmatrix}0&C\\C^{\mathsf T}&0\end{pmatrix},
 \label{eq:cauchy}
\end{equation}
the graph state is AME$(2h,q)$. Alternatively, use $2h-1$ distinct
points $\xi_1,\ldots,\xi_h,\eta_1,\ldots,\eta_{h-1}$, the same formula
for $j<h$, and $C_{ih}=1$. This extended construction works whenever
$q\geq2h-1$.
\end{lemma}
\begin{proof}
We first reduce the claim to a property of $C$. A balanced cut $A$
contains a set $\mathcal I$ of parties from the $\xi$ side and a set
$\mathcal J$ from the $\eta$ side, with $|\mathcal I|+|\mathcal J|=h$.
Write $\mathcal I^c$ and $\mathcal J^c$ for the remaining parties on each
side. Since $\Gamma$ has no edges within a side, ordering the rows as
$(\mathcal I,\mathcal J)$ and the columns as $(\mathcal J^c,\mathcal I^c)$
gives
\begin{equation}
 \Gamma_{A,A^c}=\begin{pmatrix}
   C_{\mathcal I,\mathcal J^c}&0\\
   0&(C_{\mathcal I^c,\mathcal J})^{\mathsf T}
 \end{pmatrix}.
\end{equation}
Both blocks are square, since $|\mathcal J^c|=h-|\mathcal J|=|\mathcal I|$.
So $\Gamma_{A,A^c}$ is invertible for every balanced cut as soon as every
square submatrix of $C$ is invertible, where an empty block counts as
invertible. We now check this property.

A $t\times t$ submatrix of the ordinary $C$ is again a Cauchy matrix.
After relabelling its points as $\xi_1,\ldots,\xi_t$ and
$\eta_1,\ldots,\eta_t$, its determinant is~\cite{macwilliamssloane}
\begin{equation}
 \det C'=(-1)^{\binom t2}
 \frac{\prod_{i<i'}(\xi_i-\xi_{i'})\prod_{j<j'}(\eta_j-\eta_{j'})}
      {\prod_{i,j}(\xi_i-\eta_j)}.
 \label{eq:cauchy-det}
\end{equation}
The points are distinct, so every factor is nonzero.

In the extended case, submatrices that avoid the last column are ordinary
Cauchy matrices. Let $M$ be a $t\times t$ submatrix that uses the last
column, with rows $\xi_1,\ldots,\xi_t$ and other columns
$\eta_1,\ldots,\eta_{t-1}$ after relabelling. Subtract row $t$ from every
other row. The last column becomes zero except for its final entry $1$,
and for $i,j<t$ the entry $(i,j)$ becomes
\begin{equation}
 \frac1{\xi_i-\eta_j}-\frac1{\xi_t-\eta_j}
   =\frac{\xi_t-\xi_i}{(\xi_i-\eta_j)(\xi_t-\eta_j)}.
\end{equation}
Expanding along the last column and taking the factors $\xi_t-\xi_i$ out
of the rows and $(\xi_t-\eta_j)^{-1}$ out of the columns gives
\begin{equation}
 \det M=\prod_{i<t}(\xi_t-\xi_i)\prod_{j<t}(\xi_t-\eta_j)^{-1}\,\det C',
\end{equation}
where $C'$ is the ordinary Cauchy matrix on $\xi_1,\ldots,\xi_{t-1}$ and
$\eta_1,\ldots,\eta_{t-1}$. This is nonzero by Eq.~\eqref{eq:cauchy-det}.
For $t=1$, $M=(1)$.
\end{proof}

The ordinary construction uses $N+1$ distinct field elements, so it
requires $q\geq N+1$. For $q=p^e$ with $e>1$, these must be elements of
$\F_q$, not integers reduced modulo $p$. We also need $K\leq q$ to select
$K$ projected states. Together these conditions give the claimed ranges
for $N=13,17$ and the part $q\geq17$ for $N=15$. For $N=11$, the extended
construction also covers $q=11$: take $\xi$ points $0,\ldots,5$, $\eta$
points $6,\ldots,10$, and the all-ones last column. To reach the remaining
small values of $q$ in Table~\ref{tab:families}, we use the following parents.

\subsection{Parents at smaller fields}\label{subsec:finiteparents}
The remaining fields need individual parents. We first explain how a
code supplies a parent, then record the ingredients in
Table~\ref{tab:families} and give the finite constructions.

Several parents for small $q$ come from codes over $\F_{q^2}$, in the same
way as the qubit groups come from codes over $\F_4$. In the field basis,
for $\epsilon,y\in\F_q$, the generalized Pauli operators act as
\begin{equation}
 X(\epsilon)\ket{y}=\ket{y+\epsilon},\qquad
 Z(\epsilon)\ket{y}=\chi(\epsilon y)\ket{y}.
 \label{eq:field-pauli}
\end{equation}
Up to phase, a Pauli word on $2h$ such systems is labelled by its two
parameters $\epsilon$ at each site. Combining these two elements of $\F_q$
into one element of $\F_{q^2}$ turns the word into a vector in
$\F_{q^2}^{2h}$. The larger field only keeps track of the two parameters and
adds no physical subsystem. For vectors $x,y$ let $\bar y_i=y_i^q$ and
$s=\sum_ix_i\bar y_i$. A code of dimension $h$ is \emph{Hermitian
self-dual} if $s=0$ for every pair of its vectors. Its Pauli words then
commute and fix a single state~\cite[Corollary~19]{ketkar2006}. The code is \emph{maximum distance
separable} (MDS) if every nonzero vector has at least $h+1$ nonzero
entries, the most allowed at dimension $h$. Every nonidentity word of the
group then acts on more than $h$ sites. The argument given for
Eq.~\eqref{eq:projector} therefore makes every $h$-site reduction maximally
mixed, so the state is AME$(2h,q)$~\cite{hubergrassl}.
For the parents used below, we give graph representatives and
check their balanced cuts directly. General stabilizer-to-graph
constructions are discussed in Ref.~\cite{gkr2002}.

Such a code has a generator matrix $[\Id_h\mid M]$. It is Hermitian
self-dual when $M\bar M^{\mathsf T}=-\Id_h$, and MDS when every square
submatrix of $M$ is invertible~\cite{macwilliamssloane}. These are the two exact checks made for
each parent of this kind below. The graph form then allows a separate check
of every balanced cut.

\begin{table*}[t]
\caption{Ingredients for Theorem~\ref{thm:main}, with $d=2q$ and $q$ an odd
prime power. $K$ is the number of branches. The qubit ingredient supplies
the qubit states $\ket{u_a}$ (Appendix~\ref{subsec:binary}), and a parent
supplies the $q$-level states $\ket{v_a}$. For $q\geq N+1$ the parent is a
Cauchy graph state (Lemma~\ref{lem:cauchy}). The last column lists the
parents used for smaller $q$, described below.
RS stands for Reed--Solomon. A code over $\F_{q^2}$ there still describes a
state of $q$-level systems.}
\label{tab:families}
\begin{ruledtabular}
\begin{tabular}{rcclp{8.0cm}}
$N$ & $K$ & Range & Qubit ingredient & Parents for small $q$\\
\hline
9 & 4 & $q\geq5$ & Atlas branches & $q=5$: Atlas, $q=7$: $\F_{49}$ code, $q=9$: extended RS\\
11 & 2 & $q\geq5$ & Kitamura branches & $q=5$: Bevins--Bidav, $q=7$: $\F_{49}$ code, $q=9$: explicit $\F_{81}$ code, $q=11$: extended Cauchy\\
13 & 16 & $q\geq17$ & Restricted length-$15$ code & None needed\\
15 & 8 & $q\geq9$ & Cyclic length-$15$ code & $q=9$: Hontarenko--\.{Z}yczkowski finite parent, $q=11$: $\F_{121}$ code, $q=13$: their circle parent\\
17 & 64 & $q\geq67$ & Restricted length-$18$ code & None needed
\end{tabular}
\end{ruledtabular}
\end{table*}

\paragraph*{The $q=5,7$ parents.}
For $N=9$ and $q=5$ we use Atlas's ten-party parent graph. For $N=11$ and
$q=5$ we use the AME$(12,5)$ state of Bevins and Bidav, as Kitamura
does~\cite{atlas,kitamura,bb}. For $q=7$ the parents come from Hermitian
self-dual MDS codes $[10,5,6]_{49}$ and $[12,6,7]_{49}$. The generator
matrices are $[\Id_h\mid M]$ with $h=5,6$, where $M$ is a circulant matrix
with an added first row and column. The entries of $M$ lie in
$\F_{49}=\{a+b\tau : a,b\in\F_7\}$ with $\tau^2=3$. This is a field because
$3$ is not a square modulo $7$, and its conjugation is
$\overline{a+b\tau}=a-b\tau$. The entries and graph forms are in the companion
repository. For both codes, exact arithmetic checks
$M\bar M^{\mathsf T}=-\Id_h$, every square submatrix of $M$, and every
balanced cut of the graph.

\paragraph*{The $q=9$ parents.}
For $N=9$ and $q=9$, we use a code directly over $\F_9$, rather than
a Hermitian self-dual code over $\F_{81}$. Take the polynomials of degree at most four over
$\F_9$ and record each one by its values at the nine field elements and its
degree-four coefficient. This gives an extended Reed--Solomon (RS) code
$[10,5,6]_9$. Any five entries of a codeword determine the polynomial, and
hence the other five entries. The normalized equal superposition of all
codewords is therefore AME$(10,9)$~\cite[Theorem~2]{helwigcui}, and local
Fourier transforms bring it to graph form.

For $N=11$ and $q=9$, AME$(12,9)$ is listed in the Huber--Wyderka
table~\cite{ame_table}, from the codes of Grassl and
R\"otteler~\cite{grasslroetteler}. Theorem~S7 of Hontarenko and
\.{Z}yczkowski also gives AME$(q+3,q)$ for odd prime powers $q$,
including AME$(12,9)$~\cite[Sec.~S11]{hz}. We use our own explicit code over
$\F_{81}$ with the same parameters, found by computer search, which need
not coincide with theirs. For $N=15$ and $q=9$ we use their matrix
$M_{16,9}$, which gives a Hermitian self-dual MDS code $[16,8,9]_{81}$. Its
entries are in their Eq.~(S66), the self-duality identity in Eq.~(S75), and
the check of all square submatrices in Table~S2~\cite[Sec.~S9]{hz}. The
eight branches needed at $N=15$ fit within $q=9$.

\paragraph*{The $q=11,13$ parents for fifteen parties.}
The remaining inputs for the fifteen-party family are AME$(16,11)$ and
AME$(16,13)$ parents. For $p=11$, we give the matrix of a Hermitian
self-dual MDS code explicitly. Its entries lie in
$\F_{121}=\{a+b\theta : a,b\in\F_{11}\}$ with $\theta^2=2$. This is a field
because $2$ is not a square modulo $11$, and its conjugation is
$\overline{a+b\theta}=a-b\theta$. Let $\mathbf1_7$ be the all-ones column
and define
\begin{equation}
 \begin{gathered}
 M=\begin{pmatrix}a_*&\mathbf1_7^{\mathsf T}\\
                   \mathbf1_7&M_{\rm circ}\end{pmatrix},
 \qquad a_*=8+6\theta,\\
 (M_{\rm circ})_{ij}=\boldsymbol k_{(j-i)\bmod7},\quad 0\leq i,j<7,
 \end{gathered}\label{eq:parent11}
\end{equation}
where the first row of $M_{\rm circ}$ is
\begin{equation}
 \begin{split}
 \boldsymbol k={}&(2+9\theta,\,2+4\theta,\,\theta,\,2,\\
                &10+7\theta,\,8+4\theta,\,1+3\theta).
 \end{split}\label{eq:kernel11}
\end{equation}
Exact arithmetic gives $M\bar M^{\mathsf T}=-\Id_8$ and shows that all
$12{,}869$ square submatrices of $M$ are invertible. So $M$ gives a
Hermitian self-dual MDS code $[16,8,9]_{121}$. Its graph form passes the
check for all $12{,}870$ choices of eight parties. AME$(16,11)$ is
already listed in the Huber--Wyderka table~\cite{ame_table}, which cites
Grassl and R\"otteler~\cite{grasslroetteler}. What we add is the
explicit matrix and its exact check.

For $p=13$, the circle construction of Hontarenko and \.{Z}yczkowski
(Theorem~S7, Eq.~(S81)) gives a Hermitian self-dual MDS code
$[16,8,9]_{169}$ and hence AME$(16,13)$~\cite[Sec.~S11]{hz}. The companion
repository rebuilds its matrix, converts it to a graph state over $\F_{13}$
and checks every balanced cut. Existence of AME$(16,13)$ was already
recorded by Grassl and R\"otteler~\cite[Fig.~2 of arXiv v1]{grasslroetteler}.

\subsection{From the two graphs to the coupled state}\label{subsec:assembly}
Appendix~\ref{subsec:graphbranches} specifies the qubit graph $B$ and its
ordered labels $\bl_a$. We now specify the other half of each branch and
then combine the two inputs to prove the family theorem.

\paragraph*{Reconstructing the branches.}
We can now write the $q$-level branches explicitly. Let the parent be the
graph state of $\Gamma$ on parties $0,1,\ldots,N$, and choose distinct
reference values $c_0,\ldots,c_{K-1}\in\F_q$, ordered to pair with the
qubit labels $\bl_a$. Projecting the reference onto $\ket{c_a}$ gives the
normalized state $\ket{v_a}$ of Lemma~\ref{lem:parent}. For
$y=(y_1,\ldots,y_N)\in\F_q^N$, its amplitude
$v_a(y)=\braket{y}{v_a}$ is
\begin{equation}
 v_a(y)=q^{-N/2}\chi\!\left(
    \sum_{1\leq i<j\leq N}\Gamma_{ij}y_i y_j
       +c_a\sum_{i=1}^N\Gamma_{0i}y_i\right).
 \label{eq:field-amplitude}
\end{equation}
Equations~\eqref{eq:binary-amplitude}, \eqref{eq:field-amplitude} and
\eqref{eq:coupled} thus give each state from four pieces of data: the qubit
graph $B$ with its ordered labels $\bl_a$, and the parent graph $\Gamma$
with its ordered reference values $c_a$. The companion repository lists
exactly these data, including the basis choice $\iota$ for $e>1$. One
amplitude of $\ket{\Psi}$ is a sum of $K$ terms, so the full vector of size
$(2q)^N$ never needs to be stored. The order in which the branches are
paired does not affect the AME property, but it can change the symmetries
of the state (Sec.~\ref{sec:symmetry}).

\begin{proof}[Proof of Theorem~\ref{thm:main}]
The qubit ingredients have $K=4,2,16,8,64$ branches at
$N=9,11,13,15,17$. Lemma~\ref{lem:cauchy} gives a parent whenever
$q\geq N+1$. For $N=9$ and $11$, the parents above cover $q=5,7,9$, and the
extended Cauchy parent covers $N=11$, $q=11$. For $N=15$ they cover
$q=9,11,13$, and Lemma~\ref{lem:cauchy} takes over at $q\geq17$. For $N=13$
the parent needs $q\geq14$ and the branch count needs $q\geq K=16$, so the
first odd prime power is $17$. For $N=17$ the conditions $q\geq18$ and
$q\geq64$ give $67$. In every case $K\leq q$, so we can select $K$ of the
projected states $\ket{v_c}$ of Lemma~\ref{lem:parent} and apply
Proposition~\ref{prop:coupling} with equal weights $\lambda_a=1/K$.
\end{proof}

\subsection{Exact finite data and reconstruction}\label{subsec:verification}
The finite inputs can be reconstructed separately from the coupled state.
This lets us check each ingredient before applying the analytic arguments.

The infinite families rest on the determinant and eigenspace arguments in
the text. Exact checks establish the finite inputs used by those
arguments, and numerical values serve only as a cross-check. The
companion repository~\cite{repo} supplies the ordered graphs and branch
labels, together with certificates over finite fields, the rationals and
cyclotomic numbers. Its runner records the software environment and the
result of each check. The qubit data are in
\file{data/o3_external_spec.json} and
\file{data/small_families_spec.json}, and the $q=9$ inputs are in
\file{data/prime_powers/}. The $p=11$ parent is in
\file{data/parent11_certificate.json}. The cyclic example is in the
\file{family_witness} entry of \file{data/sym1_external_spec.json}.

The external AME$(9,6)$ check has a separate input requirement, described
in Appendix~\ref{subsec:external}. Appendix~\ref{app:cyclic} explains the
cyclic witness and its exact choice of pairing.

\section{Nine-party inputs for the integer classification}\label{app:nine}
The product proof in Sec.~\ref{sec:nine} needs AME states at prime-power
dimensions and the external dimension-six input. We supply the former
first, then explain how the latter differs from our coupling.

\subsection{Every prime-power dimension at least three}\label{subsec:ninepowers}
We use the graph and field conventions of Appendix~\ref{subsec:fieldgraphs}.
Here $q=p^e$ may be even. The same character formula applies when $p=2$.

For $q=3,4,5,7,8,9$ the companion repository gives nine-party graph states
whose blocks $\Gamma_{A,A^c}$, of size $4\times5$, have rank four for every
set $A$ of four parties, as exact arithmetic confirms. The argument of
Sec.~\ref{subsec:parents} also works for such rectangular blocks: when the
rank equals the number of rows, every nonzero difference $y_A-y'_A$ still
gives a vanishing character sum over the other five parties. So these graph
states are AME$(9,q)$. The same parameters for $q=3,4,5,7,8$ also follow
from the codes of Grassl and R\"otteler~\cite[Corollary~4 and Secs.~VI.B--VI.F of arXiv v1]{grasslroetteler}.

For prime powers $q\geq8$, take the polynomials of degree at most three
over $\F_q$ and record each one by its values at nine distinct points of
$\F_q$. For $q=8$ only eight points exist, and the ninth entry is the
degree-three coefficient. This gives an RS code $[9,4,6]_q$, extended when
$q=8$. Any four entries of a codeword determine the polynomial, so every
set of four parties sees each value exactly once, and the four-party
diagonal of the equal superposition of codewords is uniform. Codewords that
differ on four parties also differ on the other five, so the off-diagonal
terms vanish after tracing. This is the code construction of
Ref.~\cite[Theorem~2]{helwigcui}, see also Ref.~\cite{goyeneche2014}, and
it proves the remaining prime-power cases.

These inputs establish the prime-power premise used in the main-text
product argument. The separate dimension-six input cannot be obtained
by taking a dimension-two AME factor.

\subsection{The external input and its verification scope}\label{subsec:external}

The nine-party classification relies on two external states, Kitamura's
AME$(9,6)$ and Atlas's AME$(9,10)$, the $q=5$ member of the $N=9$ family.
We checked both in exact arithmetic. Kitamura's state splits each six-level party
into a qubit and a three-level system as in Eq.~\eqref{eq:digits} with
$q=3$. Hontarenko and \.{Z}yczkowski describe its amplitudes, and those of the
AME$(7,6)$ of Ref.~\cite{shi7}, through complex Hadamard
blocks~\cite[Secs.~7.2--7.4 and Apps.~D--E]{hzlatin}. The companion repository does not
include Kitamura's original matrix, so without it the runner reports the
AME$(9,6)$ check as not run rather than passed. Atlas and Kitamura also give
formal proofs in the Lean proof assistant. We read fixed versions of their
sources, including Atlas's final existence theorem and Kitamura's coupling
identity and reported build status, but did not rebuild these projects.

The rank bound in Sec.~\ref{sec:limits} rules out all-cut erasure at
$q=3$, so Kitamura's state must cancel its interference differently. Its
branch decomposition shows how.
Grouped by four of its binary digits, Kitamura's state has sixteen
branches, each an AME$(9,3)$ graph state. Since sixteen exceeds three,
its interference cannot cancel in the $q$-level factor alone on every
cut, and the two factors share the cancellation, as in Ref.~\cite{shi7}.

\section{Constructing the compensated cyclic symmetry}\label{app:cyclic}
We turn the freedom in the parent branches into the matching construction
used in Sec.~\ref{sec:symmetry}. Here $N=2k+1$ and $q=p$ is prime.
The branch count is $K\leq p$, and the qubit branches are orthonormal.

\subsection{Superpositions and matching}\label{subsec:matching}
A corrected party permutation acts within a preserved branch span as a
unitary matrix, its logical action. Preserving both spans is necessary
for this description, but the actions must also agree through the pairing.

Equation~\eqref{eq:erasure} allows more than selecting individual parent
branches. Any orthonormal set of their superpositions satisfies the same
condition. To describe this freedom, let $e_a$ denote standard basis vectors. Define
$U:\mathbb C^K\to(\mathbb C^2)^{\otimes N}$ by $Ue_a=\ket{u_a}$, for
orthonormal qubit states whose average $UU^\dagger/K$ is $k$-uniform.
Define $T:\mathbb C^p\to(\mathbb C^p)^{\otimes N}$ by $Te_c=\ket{v_c}$,
$c\in\F_p$, using all $p$ projected parent states. Instead of pairing
$\ket{u_a}$ with a single $\ket{v_c}$, we may pair it with $TVe_a$ for any
isometry $V:\mathbb C^K\to\mathbb C^p$, that is, any $V$ with
$V^\dagger V=\Id$. The state
\begin{equation}
 \ket{\Psi_V}=K^{-1/2}\sum_{a=0}^{K-1}Ue_a\otimes TVe_a
 \label{eq:isometry}
\end{equation}
is AME$(N,2p)$ for every such $V$. To see this, expand any
$x,y\in\mathbb C^p$ in the reference basis and apply
Eq.~\eqref{eq:erasure} term by term. This gives
\begin{equation}
 \Tr_{A^c}T\ketbra{x}{y}T^\dagger
       =\braket{y}{x}\Id_A/p^k.
 \label{eq:encoding}
\end{equation}
The vectors $Ve_a$ are orthonormal, so the states $TVe_a$ satisfy
Eq.~\eqref{eq:erasure} and Proposition~\ref{prop:coupling} applies.

Now suppose the same party permutation, with local corrections on each
factor, preserves both branch spans. Write its logical actions as
$\alpha$ on the qubits and $\beta$ on the $p$-level states. We seek an
isometry $V$ for which these actions preserve the pairing. With
$\overline\alpha$ the entrywise complex conjugate and $\vartheta$ a real
phase, the required condition is
\begin{equation}
 \beta V=e^{i\vartheta}V\overline\alpha.
 \label{eq:matching}
\end{equation}
Indeed, in the two branch bases the coefficient matrix of $\ket{\Psi_V}$
is proportional to $V^{\mathsf T}$, and that of the transformed state to
$\alpha(\beta V)^{\mathsf T}$.

For the cyclic family below, the party permutation has order fifteen,
while both logical actions have order three. This suffices because
Eq.~\eqref{eq:matching} constrains only the logical actions.

\subsection{The fifteen-party construction}\label{subsec:cyclicproof}
We now prove Theorem~\ref{thm:c15}. The qubit input is the length-15
construction of Appendix~\ref{subsec:codes}, with its eight branches in
the ordered graph basis of Appendix~\ref{subsec:graphbranches}.

\begin{proof}
We construct the same cyclic shift on both factors, then choose $V$ to
match their logical actions. The qubit action is fixed by our eight
branches. We choose a parent whose logical action has enough eigenvectors
with each required eigenvalue.

The length-15 qubit code is cyclic, so it is unchanged by the cyclic
shift of the fifteen parties. In the graph basis of the branches, the
shift is followed by local Clifford operations and a Pauli correction that
restores the signs of the group. The logical action $\alpha$ of this
operation on the eight branches has order three. The exact operation and
its matrix are recorded in the companion repository. In the ordered branch
basis the action is
\begin{equation}
 \begin{gathered}
 e_0\mapsto e_4\mapsto e_5\mapsto e_0,\qquad e_1,e_2\ \text{fixed},\\
 e_3\mapsto-e_7,\qquad e_7\mapsto e_6,\qquad e_6\mapsto-e_3.
 \end{gathered}\label{eq:binary-cycle}
\end{equation}
An exact check confirms that the physical operation induces this action.
Each signed three-cycle has sign product $+1$, so $\alpha^3=\Id$. With
$\zeta=e^{2\pi i/3}$, the eigenspaces of $\alpha$ for $1,\zeta,\zeta^2$
have dimensions $(4,2,2)$.

For the $p$-level factor, we need a parent that admits the same cyclic
shift of fifteen parties. Choose $g\in\F_p$ of multiplicative order
fifteen, which exists because $15\mid p-1$. Take the polynomials $f$ of degree at most seven over $\F_p$
and record each one by its values at $1,g,\ldots,g^{14}$ together with its
degree-seven coefficient, which is the entry of the reference party. This
is an extended RS code $[16,8,9]_p$. As for $q=9$ above, any eight entries
determine $f$, so the equal superposition of codewords is
AME$(16,p)$~\cite[Theorem~2]{helwigcui}.

Replacing $f(z)$ by $f(gz)$ maps the code to itself. It shifts the fifteen
evaluation entries cyclically and multiplies the reference entry by $g^7$.
Multiplying every entry of every codeword by the same nonzero $\gamma$ also
maps the code to itself. Let $\mu\in\F_p$ have order three and take
$\gamma=\mu g^{-7}$. The combined operation shifts the fifteen parties
cyclically and multiplies the reference value by $\mu$. Multiplication by a
field element is a local Clifford operation, and it stays one after the
local Fourier transforms to graph form. So the $p$-level side has the same
cyclic shift of the parties, with logical action $\Pi\ket{c}=\ket{\mu c}$
on the projected states, $c\in\F_p$.

The permutation $\Pi$ fixes $0$ and splits the nonzero labels into
$(p-1)/3$ cycles of length three. Its eigenspaces for $1,\zeta,\zeta^2$
therefore have dimensions
\begin{equation}
 \left(\frac{p+2}{3},\frac{p-1}{3},\frac{p-1}{3}\right).
 \label{eq:prime-cycle}
\end{equation}
Since $p\geq31$, these are at least $(11,10,10)$, larger than the
dimensions $(4,2,2)$ of the eigenspaces of $\overline\alpha$. Mapping each
eigenspace of $\overline\alpha$ isometrically into the eigenspace of $\Pi$
with the same eigenvalue gives an isometry $V$ with
$\Pi V=V\overline\alpha$. Equation~\eqref{eq:isometry} gives AME, and
Eq.~\eqref{eq:matching} with $\beta=\Pi$ gives the symmetry.
\end{proof}

\subsection{A definite choice of pairing}\label{subsec:cyclicpairing}

To fix $V$ concretely, form the projectors onto the eigenspaces of
$\overline\alpha$ and of $\Pi$, and apply Gram--Schmidt to their columns in
the standard order, dropping zero vectors. Map the resulting basis of each
eigenspace of $\overline\alpha$ to the first equally many basis vectors of
the eigenspace of $\Pi$ with the same eigenvalue. The companion repository
carries this out in exact cyclotomic arithmetic. The $p=31$ example uses
$\mu=5$.

\section{Capacity and arbitrary-spectrum rank bounds}\label{app:rank}\label{app:limits}
The comparison in Sec.~\ref{sec:limits} needs an upper bound on the
number of erasing branches and lower bounds on the rank of the qubit
mixture. We prove them separately so that neither depends on a projector
assumption.

\subsection{At most q erasing branches}\label{subsec:capacity}
Let $N=2k+1$ and let $q\geq2$ be any integer.

First we bound the number of $q$-level branches, without assuming that
they come from a parent. The argument is the entropic proof of the
quantum Singleton bound due to Cerf and Cleve~\cite{cerfcleve1997}, see
also Ref.~\cite{grasslhuberwinter}. Suppose $K$ states $\ket{v_a}$ satisfy
Eq.~\eqref{eq:erasure}, and record their label in a reference system
$\mathrm{Ref}$ of dimension $K$. The resulting pure state is
$K^{-1/2}\sum_a\ket{a}_{\mathrm{Ref}}\ket{v_a}$. Split the $N=2k+1$
parties into $A_1,A_2,A_3$ of sizes $k,k,1$. Equation~\eqref{eq:erasure}
makes the reduced state on $\mathrm{Ref}\,A_j$ equal to
$\Id/K\otimes\Id/q^k$ for $j=1,2$. Let $S_{\rm vN}$ denote von Neumann
entropy. Because these reductions are products,
$S_{\rm vN}(\mathrm{Ref}\,A_1)=S_{\rm vN}(\mathrm{Ref})+S_{\rm vN}(A_1)$.
Purity of the full state gives
$S_{\rm vN}(\mathrm{Ref}\,A_1)=S_{\rm vN}(A_2A_3)$, which subadditivity
bounds by $S_{\rm vN}(A_2)+S_{\rm vN}(A_3)$. The same argument with $A_1$
and $A_2$ exchanged gives the second line of
\begin{align}
 S_{\rm vN}(\mathrm{Ref})+S_{\rm vN}(A_1)
    &\leq S_{\rm vN}(A_2)+S_{\rm vN}(A_3),\notag\\
 S_{\rm vN}(\mathrm{Ref})+S_{\rm vN}(A_2)
    &\leq S_{\rm vN}(A_1)+S_{\rm vN}(A_3).
 \label{eq:capacity-proof}
\end{align}
Adding them gives the quantum Singleton
bound~\cite[Theorem~2]{rains1999nonbinary}, see also Ref.~\cite{hubergrassl},
\begin{equation}
 \log K=S_{\rm vN}(\mathrm{Ref})\leq S_{\rm vN}(A_3)\leq\log q,
 \qquad K\leq q.
 \label{eq:capacity}
\end{equation}
This holds for any states satisfying Eq.~\eqref{eq:erasure}, not only for those obtained from a parent.

\subsection{Necessary inequalities for the qubit mixture}\label{subsec:rankconstraints}

To prove Proposition~\ref{prop:rank}, we assume a rank bound and derive
linear inequalities for sums of Pauli correlations. We then combine the
inequalities to obtain a contradiction. No restriction on the spectrum
is needed. Let $\mathcal P_N=\{I,X,Y,Z\}^{\otimes N}$ without scalar
phases. For a density matrix $\rho$ and $j=0,\ldots,N$, define
\begin{align}
 \mathfrak a_j&=\sum_{\substack{E\in\mathcal P_N\\\wt(E)=j}}
                      (\Tr E\rho)^2,\notag\\
 \mathfrak b_j&=\sum_{\substack{E\in\mathcal P_N\\\wt(E)=j}}
                      \Tr\ab(E\rho E\rho).
 \label{eq:enumerators}
\end{align}
Collect these sums into polynomials in formal variables $z,w$:
$\mathcal E_a(z,w)=\sum_j\mathfrak a_jz^{N-j}w^j$ and
$\mathcal E_b(z,w)=\sum_j\mathfrak b_jz^{N-j}w^j$. These are the
unnormalized Shor--Laflamme weight enumerators, in the convention of
Rains~\cite{shorlaflamme1997,rains1998we}, evaluated on $\rho$.
The MacWilliams transform relates them. A second transform, called the shadow transform,
defines $\mathcal E_s(z,w)=\sum_j\mathfrak s_jz^{N-j}w^j$ and its
nonnegative coefficients $\mathfrak s_j$:
\begin{align}
 \mathcal E_b(z,w)&=2^{-N}\mathcal E_a(z+3w,z-w),\notag\\
 \mathcal E_s(z,w)&=2^{-N}\mathcal E_a(z+3w,w-z),
 \qquad\mathfrak s_j\geq0.
 \label{eq:transforms}
\end{align}
These identities and positivity hold for positive operators, not only
projectors~\cite[Theorems~1,6,8 and Corollary~7]{rains}. Their Pauli-correlation
formulation is discussed in Ref.~\cite{shadowbounds}.

For a positive integer $R$, suppose $\rank\rho\leq R$. Hilbert--Schmidt
Cauchy--Schwarz applied to $\rho^{1/2}E\rho^{1/2}$ on its support gives
$(\Tr E\rho)^2\leq R\Tr\ab(E\rho E\rho)$, hence
\begin{equation}
 \mathfrak a_j\leq R\mathfrak b_j.
 \label{eq:rank-pauli}
\end{equation}
For a normalized projector of rank $R$ this is an inequality of
Rains~\cite[Theorem~2]{rains1998we}. Here the spectrum is arbitrary.
A second rank constraint compares the purities of complementary
reductions. First express these purities using the same Pauli sums.
For an integer $s$ between zero and $N$, Pauli expansion on $A$ gives
\begin{align}
 \sum_{|A|=s}\Tr\rho_A^2
  &=2^{-s}\sum_{j=0}^s\binom{N-j}{s-j}\mathfrak a_j,\notag\\
 \sum_{|A|=s}\Tr\rho_{A^c}^2
  &=2^{-s}\sum_{j=0}^s\binom{N-j}{s-j}\mathfrak b_j.
 \label{eq:purity-expansions}
\end{align}
The binomial factor counts the size-$s$ subsets containing a given
weight-$j$ Pauli support. For the second identity, summing conjugations
by all Paulis supported on $A$ replaces that factor by the identity
and leaves the complementary reduction.

Now purify $\rho$ into a reference of dimension at most $R$. The
partial-trace Hilbert--Schmidt bound on that reference gives
$\Tr\rho_{A^c}^2\leq R\Tr\rho_A^2$. Summing over $A$ and cancelling
the common factor $2^{-s}$ yields
\begin{equation}
 \sum_{j=0}^s\binom{N-j}{s-j}\mathfrak b_j
 \leq R\sum_{j=0}^s\binom{N-j}{s-j}\mathfrak a_j.
 \label{eq:complement-purity}
\end{equation}
The other necessary conditions are $\mathfrak a_j\geq0$,
$\mathfrak a_0=1$, $\mathfrak a_1=\cdots=\mathfrak a_k=0$ for
$k$-uniformity, $\mathfrak b_0=\Tr\rho^2\leq1$, and
\begin{equation}
 \sum_{j=0}^s\binom{N-j}{s-j}\mathfrak a_j\leq\binom Ns2^s,
 \qquad 0\leq s\leq N.
 \label{eq:purity-upper}
\end{equation}
Equation~\eqref{eq:purity-upper} follows because every reduced density
matrix has purity at most one. After substituting the transforms, all
these conditions are linear inequalities with rational coefficients in
the $\mathfrak a_j$. We exclude a proposed rank $R$ by finding a
nonnegative combination that gives a negative constant. No equality
$\Tr\rho^2=1/K$ has been assumed, so the argument covers unequal
eigenvalues as required by the general coupling.

\subsection{Exact contradictions and their scope}\label{subsec:rankproof}
The inequalities now let us exclude a proposed rank by an exact
nonnegative combination. We first summarize the four cases, then display
the nine- and thirteen-qubit identities and identify the other certificates.

\begin{proof}[Proof of Proposition~\ref{prop:rank}]
The preceding subsection gives linear inequalities that every $k$-uniform
state of rank at most $R$ must satisfy. A nonnegative combination of
these inequalities must remain nonnegative. For $N=13$, $R=8$, the
identity below displays a combination that instead equals a negative constant,
excluding every rank up to eight. For $(N,R)=(9,3)$, $(15,7)$ and
$(17,22)$, exact rational combinations of the same inequalities give
$-288/5473$, $-2031120/18916487$ and $-6528/6182327$. These certificates likewise
exclude every rank up to the stated $R$.
\end{proof}

For $N=9,k=4,R=3$, set
$D_j=3\mathfrak b_j-\mathfrak a_j\geq0$ and
\begin{equation}
 D_7^{\rm c}=\sum_{j=0}^7\binom{9-j}{7-j}
                         (3\mathfrak a_j-\mathfrak b_j)\geq0.
 \label{eq:slack9}
\end{equation}
These inequalities follow from Eqs.~\eqref{eq:rank-pauli}
and~\eqref{eq:complement-purity}. Substituting the transforms,
$\mathfrak a_0=1$ and $\mathfrak a_1=\cdots=\mathfrak a_4=0$ gives
\begin{equation}
 \begin{split}
 &3456\mathfrak s_0+128D_1+1152\mathfrak s_1\\
 &\qquad+160D_2+576\mathfrak s_2+D_7^{\rm c}=-288.
 \end{split}\label{eq:farkas9}
\end{equation}
All coefficients of $\mathfrak a_5,\ldots,\mathfrak a_9$ cancel.
Every term on the left is nonnegative, which excludes rank at most
three. Dividing by $5473$ gives the normalized contradiction
$-288/5473$.

For $N=13,k=6,R=8$, define the nonnegative differences
$D_j=8\mathfrak b_j-\mathfrak a_j$ from Eq.~\eqref{eq:rank-pauli}, and
\begin{equation}
 D_{10}^{\rm c}=\sum_{j=0}^{10}\binom{13-j}{10-j}
                         (8\mathfrak a_j-\mathfrak b_j).
 \label{eq:slack}
\end{equation}
Substitute the transforms, $\mathfrak a_0=1$ and
$\mathfrak a_1=\cdots=\mathfrak a_6=0$. Then
\begin{equation}
 \begin{split}
 &292864\mathfrak s_0+8192D_1+133120\mathfrak s_1+3584D_2\\
 &\quad+39936\mathfrak s_2+1536D_3+13312\mathfrak s_3+D_{10}^{\rm c}\\
 &\hspace{3cm}=-3328.
 \end{split}\label{eq:farkas}
\end{equation}
All coefficients of $\mathfrak a_7,\ldots,\mathfrak a_{13}$ cancel.
The left side is nonnegative, so no state of rank at most eight
can meet the conditions. This is a Farkas certificate: a nonnegative
combination of necessary inequalities gives a negative constant.
Dividing by $492545$ gives the normalized contradiction
$-3328/492545$ in \file{data/o4_lp_certificates.json}.

For $(N,R)=(15,7),(17,22)$, the file
\file{data/general_rank_certificates.json} supplies nonnegative rational
multipliers for the same system. The checker
\file{coupled_ame/general_rank.py} reconstructs each inequality and
replays the multipliers exactly, producing the values in the proof
of Proposition~\ref{prop:rank}. The nine- and thirteen-qubit identities above can be checked by hand;
the remaining two cases are replayed exactly by the checker. At ranks $4,9,8,23$ the inequalities have exact solutions,
so the bounds of Proposition~\ref{prop:rank} are the best these inequalities
give. Such solutions do not show that states of those ranks exist.

\subsection{Thirteen-party consequences in either orientation}\label{subsec:consequences}
We finish by checking the two thirteen-party examples in
Sec.~\ref{sec:limits}. Here $N=13$ and $k=6$, and we use the same
weighted coupling and arbitrary-spectrum rank bound as in the main text.

For AME$(13,10)$ the coupling would need a qubit mixture of rank at most
five, which the rank-nine bound excludes. Exchanging the roles of the two
factors also fails. The qubit states would then satisfy
Eq.~\eqref{eq:erasure}, so each of them would be a pure six-uniform state of
thirteen qubits, which the same bound excludes. Neither argument rules out
other constructions of AME$(13,10)$, for example ones in which the
interference cancels differently on different cuts, as in
Ref.~\cite{shi7}. For AME$(13,26)$, a six-uniform qubit mixture of rank
between nine and thirteen would suffice. Its eigenvectors could be paired
with projected states of the AME$(14,13)$ parent from the extended Cauchy
construction. We have not found such a mixture.

\section{Coupling at general odd party counts}\label{app:general-odd}
The finite ingredients in the main text give small thresholds. We now
separate the mechanism that works for every odd party count from the
choice of those ingredients. Throughout this appendix, $N=2k+1\geq3$
with integer $k\geq1$. The arguments use elementary projector and packing
constructions. We include their proofs to make the scope explicit,
without claiming novelty for the deductions.

\subsection{The rank criterion}\label{subsec:odd-criterion}
Recall that $\mu_N$ is the minimum rank of a $k$-uniform density matrix
on $N$ qubits, without a spectrum or stabilizer assumption.

\begin{proof}[Proof of Theorem~\ref{thm:odd-rank}]
Put $h=k+1$, so $N+1=2h$. The extended construction in
Lemma~\ref{lem:cauchy} uses $2h-1=N$ distinct points of $\F_q$,
where $\F_q$ is the field of $q$ elements, and an all-ones last column.
Since $q\geq N$, it supplies an AME$(N+1,q)$ parent.

Suppose $\rho$ attains $\mu_N=R\leq q$. Its spectral decomposition
$\rho=\sum_{a=0}^{R-1}\lambda_a\ketbra{u_a}{u_a}$ has positive
probabilities $\lambda_a$ summing to one and orthonormal eigenvectors
$\ket{u_a}$. Select $R$ distinct reference labels of the parent.
Lemma~\ref{lem:parent} gives states $\ket{v_a}$ satisfying
Eq.~\eqref{eq:erasure}, while $k$-uniformity of $\rho$ gives
Eq.~\eqref{eq:binary-average}. Proposition~\ref{prop:coupling} then
gives AME$(N,2q)$.

Conversely, any such coupling with $K$ branches has a $k$-uniform
qubit mixture $\rho_{\rm bin}$ of rank at most $K$. The capacity
bound in Eq.~\eqref{eq:capacity} gives $K\leq q$, without requiring
that the branches come from a parent. Therefore
$\mu_N\leq\operatorname{rank}\rho_{\rm bin}\leq K\leq q$.
\end{proof}

\paragraph*{Remark.}
For any integer $q\geq2$, a given AME$(N+1,q)$ parent and a
$k$-uniform qubit density matrix of rank at most $q$ suffice by the
same spectral-decomposition proof. Below $q=N$, the rank inequality
alone does not provide the parent or another suitable set of branches.

\subsection{An explicit mixture for every odd N}\label{subsec:odd-mixture}
We can always meet the qubit requirement, at the cost of many branches.
Let $I,X,Y,Z$ denote the identity and the three single-qubit Pauli
matrices, and let $\Id$ denote the identity on all $N$ qubits. Set
\[
 G_X=X^{\otimes2k}\otimes I,\qquad
 G_Z=Z^{\otimes2k}\otimes I.
\]
The two operators commute because $2k$ is even. They are independent
Hermitian involutions, and every nonidentity element of the group they
generate has weight $2k=N-1$. In particular the group does not contain
$-\Id$.
For odd $N>7$, this is the construction of
K{\l}obus et al.~\cite[Sec.~V.M, Eq.~(29)]{klobus2019mixed}, up to a
permutation of the parties and a choice of Pauli sign sector: the product
of their $X$- and $Y$-type generators is $(-1)^k$ times the $Z$-type word.
The proof below also covers the smaller odd values $N\geq3$.

\begin{proof}[Proof of Corollary~\ref{cor:odd-N}]
The common fixed-space projector and its rank are
\[
 P_N=\frac{(\Id+G_X)(\Id+G_Z)}4,\qquad K=2^{N-2}.
\]
The rank follows by taking the trace, since all three nonidentity
Pauli terms are traceless. Since $N-1>k$, the weight criterion of
Appendix~\ref{subsec:weights} shows that $P_N/K$ is $k$-uniform.

For an explicit basis, let $x$ range over binary strings of length
$2k$ with first bit zero and even weight. Let $\bar x$ be the bitwise
complement and let $b\in\{0,1\}$. The vectors
\[
 \ket{u_{x,b}}=\frac{\ket{x}+\ket{\bar x}}{\sqrt2}\otimes\ket{b}
\]
are fixed by $G_X$ and $G_Z$. They are orthonormal because distinct
$x$ select disjoint complementary pairs and the final-qubit labels
are orthogonal. There are $2^{2k-2}$ allowed strings and two choices
of $b$, giving exactly $K=2^{N-2}$ vectors. Thus they form a basis of
the range of $P_N$.

For $q\geq B_N=\max\{N,2^{N-2}\}$, the extended-Cauchy parent exists
and has at least $K$ projected states. Order the pairs $(x,b)$
lexicographically and pair them with any $K$ distinct reference labels.
Use equal probabilities $1/K$ in Proposition~\ref{prop:coupling}.
The resulting state is normalized and AME$(N,2q)$.
\end{proof}

\subsection{Growth of the branch requirement}\label{subsec:odd-growth}
The sufficient bound above is large, but a polynomial branch count
cannot work asymptotically within this coupling either. A purification
makes this lower bound transparent.

We use the orthogonality counting behind the nondegenerate quantum
Hamming bound~\cite[Chap.~7.1, Eq.~(7.1)]{gottesman1997hamming} on a
purification, with errors acting only on the $N$ qubits. The required
orthogonality follows from $k$-uniformity, as the proof shows.

\begin{proposition}[Rank packing bound]\label{prop:odd-packing}
Set $t=\lfloor k/2\rfloor$ and
$V_N(t)=\sum_{j=0}^{t}3^j\binom Nj$. Every $k$-uniform density matrix
of $N$ qubits and rank $R$ satisfies
\[
 R\geq\left\lceil\frac{V_N(t)}{2^N}\right\rceil.
\]
Consequently every coupling satisfying
Proposition~\ref{prop:coupling} requires
$q\geq\lceil V_N(t)/2^N\rceil$.
\end{proposition}

\begin{proof}
Purify the density matrix $\rho$ into a unit vector $\ket{\Omega}$
on the $N$ qubits and a reference of dimension $R$. Let $E$ range over
Pauli words in $\{I,X,Y,Z\}^{\otimes N}$ with weight at most $t$,
and let $I_{\rm ref}$ be the identity on the reference. There are
$V_N(t)$ vectors $(E\otimes I_{\rm ref})\ket{\Omega}$.

Each vector is normalized. For distinct words $E,F$, the operator
$E^\dagger F$ is a nonidentity Pauli word up to phase and has weight
at most $2t\leq k$. Uniformity gives
$\operatorname{Tr}(\rho E^\dagger F)=0$, since the reduction on its
support is maximally mixed. Thus these vectors are mutually orthogonal.
They lie in dimension $2^N R$, so $V_N(t)\leq2^N R$. The rank bound
follows. Combining it with
$\operatorname{rank}\rho_{\rm bin}\leq K\leq q$ proves the coupling
bound without any spectrum or parent assumption.
\end{proof}

\paragraph*{Remark.}
Let $Q_N$ be the smallest odd prime power for which this all-cut
coupling exists, and keep $B_N=\max\{N,2^{N-2}\}$. Such a $Q_N$
exists by Corollary~\ref{cor:odd-N}. Choosing the first power of three
at least $B_N$ gives
\[
 \begin{gathered}
 \left\lceil\frac{V_N(\lfloor(N-1)/4\rfloor)}{2^N}\right\rceil
       \leq\mu_N,\\
 \mu_N\leq Q_N<3B_N.
 \end{gathered}
\]
To see the lower exponential rate, retain the last term of $V_N(t)$.
Integral bounds on the sum of $\log j$ give the factorial estimate
$\log(n!)=n\log n-n+O(\log n)$. It yields
$\log_2\binom Nt=N H_2(t/N)+O(\log N)$, where
$H_2(x)=-x\log_2x-(1-x)\log_2(1-x)$ is binary entropy.
Here $t/N\to1/4$, so
\[
 \begin{aligned}
 \liminf_{\substack{N\to\infty\\N\text{ odd}}}
       \frac{\log_2\mu_N}{N}
 &\geq H_2(1/4)+\frac14\log_2 3-1\\
 &=1-\frac12\log_2 3>0.
 \end{aligned}
\]
The same lower rate holds for $Q_N$. The upper bounds
$\mu_N\leq2^{N-2}$ and $Q_N<3B_N$ give upper rate at most one.
These estimates do not determine an optimal rate. They show that a
polynomial $q$ cannot suffice for arbitrarily large odd $N$ in this
coupling, without ruling out other AME constructions.

\paragraph*{Open question.}
Determine the minimum ranks $\mu_N$ for larger odd $N$ and their optimal
asymptotic growth. The rank inequalities give necessary conditions,
but a feasible solution is not a density matrix. Better thresholds
require actual mixtures and, at small fields, suitable $q$-level branches.

\bibliography{references}
\end{document}